\documentclass[conference]{IEEEtran}
\IEEEoverridecommandlockouts
\usepackage{cite}
\usepackage{amsmath,amssymb,amsfonts}
\usepackage{algorithmic}
\usepackage{graphicx}
\usepackage{textcomp}
\usepackage{xcolor}
\usepackage{booktabs}

\def\BibTeX{{\rm B\kern-.05em{\sc i\kern-.025em b}\kern-.08em
    T\kern-.1667em\lower.7ex\hbox{E}\kern-.125emX}}

\begin{document}
\bstctlcite{IEEEexample:BSTcontrol}

\title{An Ultra-Low-Power Synthesizable Asynchronous AER Encoder for Neuromorphic Edge Devices\\
% {\footnotesize \textsuperscript{*}Note: Sub-titles are not captured in Xplore and
% should not be used}
% \thanks{Identify applicable funding agency here. If none, delete this.}
}

\author{
\IEEEauthorblockN{Yihui Wang$^{\dagger}$, Sheng-Yu~Peng$^{*}$, Sahil~Shah$^{\dagger}$}

\IEEEauthorblockA{$^{\dagger}$University of Maryland, College Park, MD, USA\\
\{yihuiw, sshah389\}@umd.edu}

\IEEEauthorblockA{$^{*}$National Yang Ming Chiao Tung University, Taiwan\\
speng@nycu.edu.tw}
}
\maketitle

\begin{abstract}
This paper presents a fully synthesizable, tree-based Address-Event Representation (AER) encoder designed for scalable neuromorphic computing systems. To achieve high throughput while maintaining strict compatibility with commercial EDA workflows, the asynchronous design employs a bundled-data protocol within a semi-decoupled micropipeline. The architecture replaces traditional transparent latches with standard edge-triggered flip-flops, enabling digital synthesis and place-and-route (PnR) using Cadence toolkits. A cross-coupled NAND-based random-priority arbiter is embedded within the encoder of each tree node to resolve event collisions efficiently. An 8-event AER prototype is fabricated in 65 nm CMOS technology utilizing a purely digital standard-cell flow. Post-fabrication silicon measurements validate the design, demonstrating a peak throughput of 33~MEvent/s and an average event latency of 50~ns, equating to a propagation delay of 17~ns/(event-bit). The design consumes only 435~fJ per encoded event.
\end{abstract}


\section{Introduction}

Neuromorphic computing offers a highly energy-efficient, brain-inspired paradigm for artificial intelligence by relying on sparse, event-driven spikes \cite{kudithipudi_neuromorphic_2025}. However, efficiently routing these events across millions of neurons poses a critical architectural challenge. While neural traffic is predominantly sparse, it frequently features dense, localized bursts, such as a dynamic vision sensor pixel jumping from millihertz background activity to kilohertz firing rates when stimulated \cite{huo_research_2025, lichtsteiner_128times_2008}. Traditional synchronous interconnects struggle with this unpredictable traffic, as constant global clock synchronization induces severe power overhead, latency bottlenecks, and network congestion.

To address this, the Address-Event Representation (AER) protocol time-multiplexes spikes into digital addresses over shared buses \cite{park_hierarchical_2017, purohit_hierarchical_2021, zamarreno-ramos_multicasting_2013}. To further optimize these links, designers increasingly adopt the asynchronous bundled-data protocol \cite{sutherland_micropipelines_1989}. In this approach, the global clock is replaced by localized handshaking circuits that bundle data lines with control signals. Memory elements, such as flip-flops (FFs) or latches, are triggered directly by asynchronous "request" and "acknowledge" signals rather than a continuous clock edge. By capturing data only when valid events are present and remaining idle otherwise, this event-driven mechanism naturally adapts to neuromorphic traffic, sustaining high throughput with minimal active power \cite{ouyang_scalable_2024}.

Despite these benefits, integrating asynchronous logic into conventional synchronous design flows is difficult. Existing implementations often rely on labor-intensive, full-custom circuits with limited scalability \cite{qiao_bi-directional_2018}, or utilize experimental, custom EDA tools that introduce steep learning curves \cite{purohit_field-programmable_2022}. To bridge this gap, this paper presents a fully synthesizable, tree-based asynchronous AER encoder implemented in 65~nm CMOS. To ensure seamless compatibility with commercial EDA workflows, we employ a four-phase bundled-data protocol. Unlike traditional micropipelines reliant on transparent latches \cite{sutherland_micropipelines_1989}, our architecture utilizes standard edge-triggered FFs within a semi-decoupled micropipeline, allowing for straightforward logic synthesis and place-and-route. Enhanced by embedded cross-coupled NAND-based random-priority arbiters for efficient collision resolution, post-fabrication silicon measurements validate the design, demonstrating a peak throughput of 33~MEvent/s and an exceptional energy efficiency of 435~fJ/event.

\begin{figure}[htbp]
    \centering
    \includegraphics[width=\columnwidth]{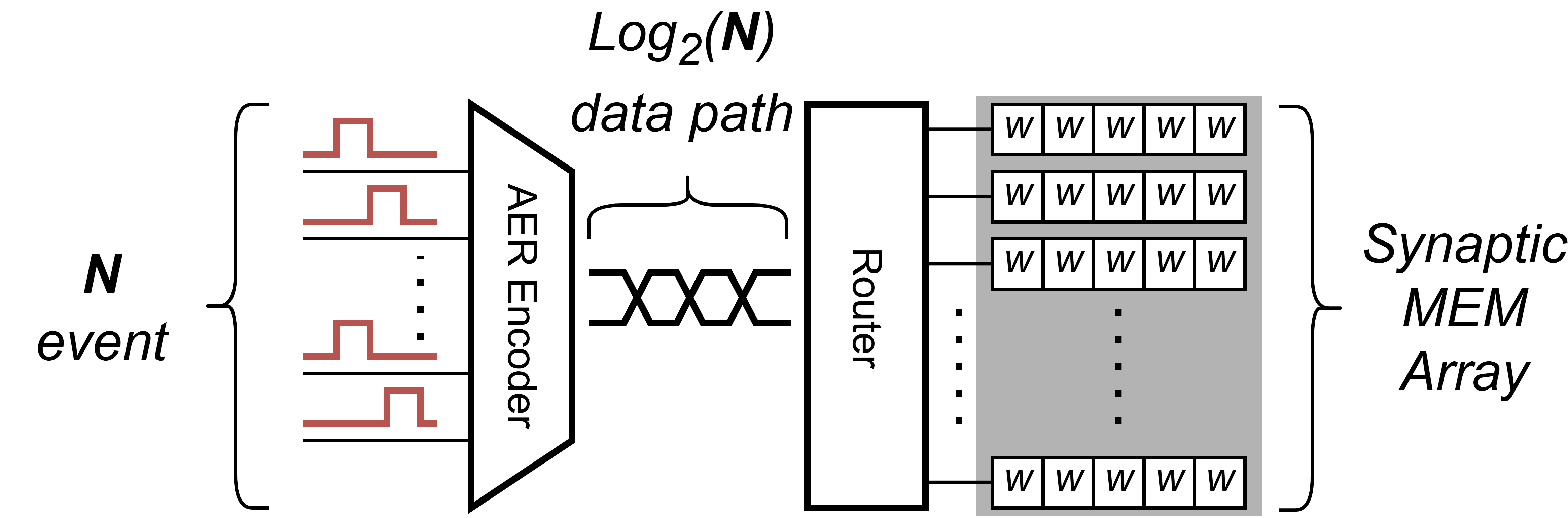}
    \caption{The Address-Event Representation (AER) protocol. The AER system encodes events from the neuron array, and the resulting address is routed for downstream synaptic memory access.}
    \label{fig:aer_protocol}
\end{figure}

\begin{figure}[htbp]
    \centering
    \includegraphics[width=0.9\columnwidth]{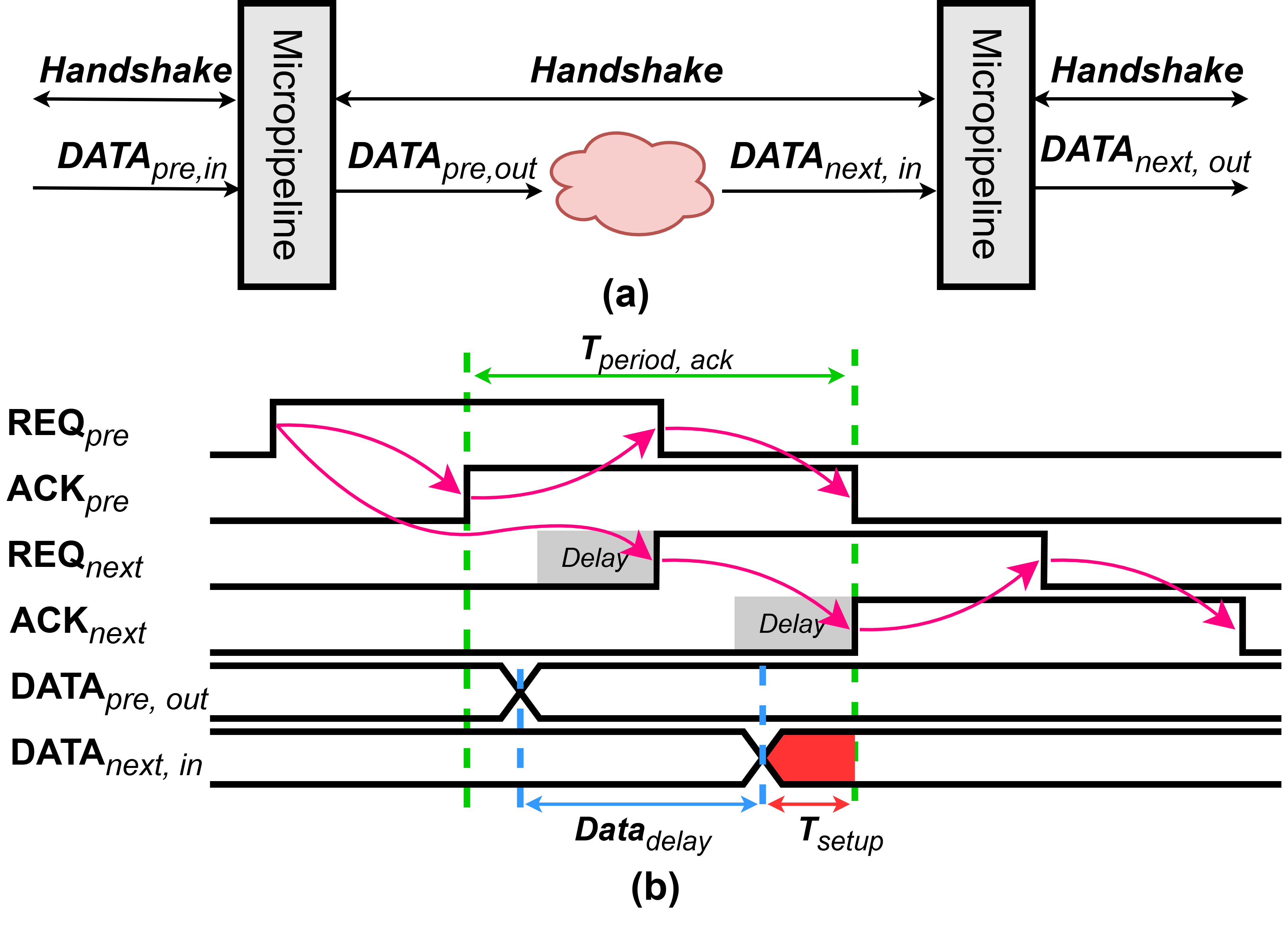}
    \caption{The proposed asynchronous bundled-data micropipeline. (a) Datapath architecture detailing the routing of handshake and data signals between adjacent stages. (b) Time sequence diagram of the modified protocol. $\mathrm{REQ}_{\mathrm{pre}}$ and $\mathrm{ACK}_{\mathrm{pre}}$ interface with the previous stage; $\mathrm{REQ}_{\mathrm{next}}$ and $\mathrm{ACK}_{\mathrm{next}}$ interface with the next.}
    \label{fig:handshake_timing}
\end{figure}

\section{Proposed Asynchronous Architecture}

\subsection{Address-Event Representation (AER) Protocol}

To facilitate efficient, event-driven communication without the severe power overhead of global clocks, the neuromorphic SoCs employ localized asynchronous handshakes via the Address-Event Representation (AER) protocol \cite{zhong_paicore_2025}. Mimicking biological neural networks, discrete neural spikes are usually packetized into spatial digital addresses and routed directly to destination synaptic memories \cite{frenkel_morphic_2019}. Fig.~\ref{fig:aer_protocol} shows the block diagram of the AER encoder and routing to synaptic memory. Valid address data is transmitted using a localized Request (REQ) and Acknowledge (ACK) handshake (Fig.~\ref{fig:handshake_timing}).

\begin{figure}[htbp]
    \centering
    \includegraphics[width=\columnwidth]{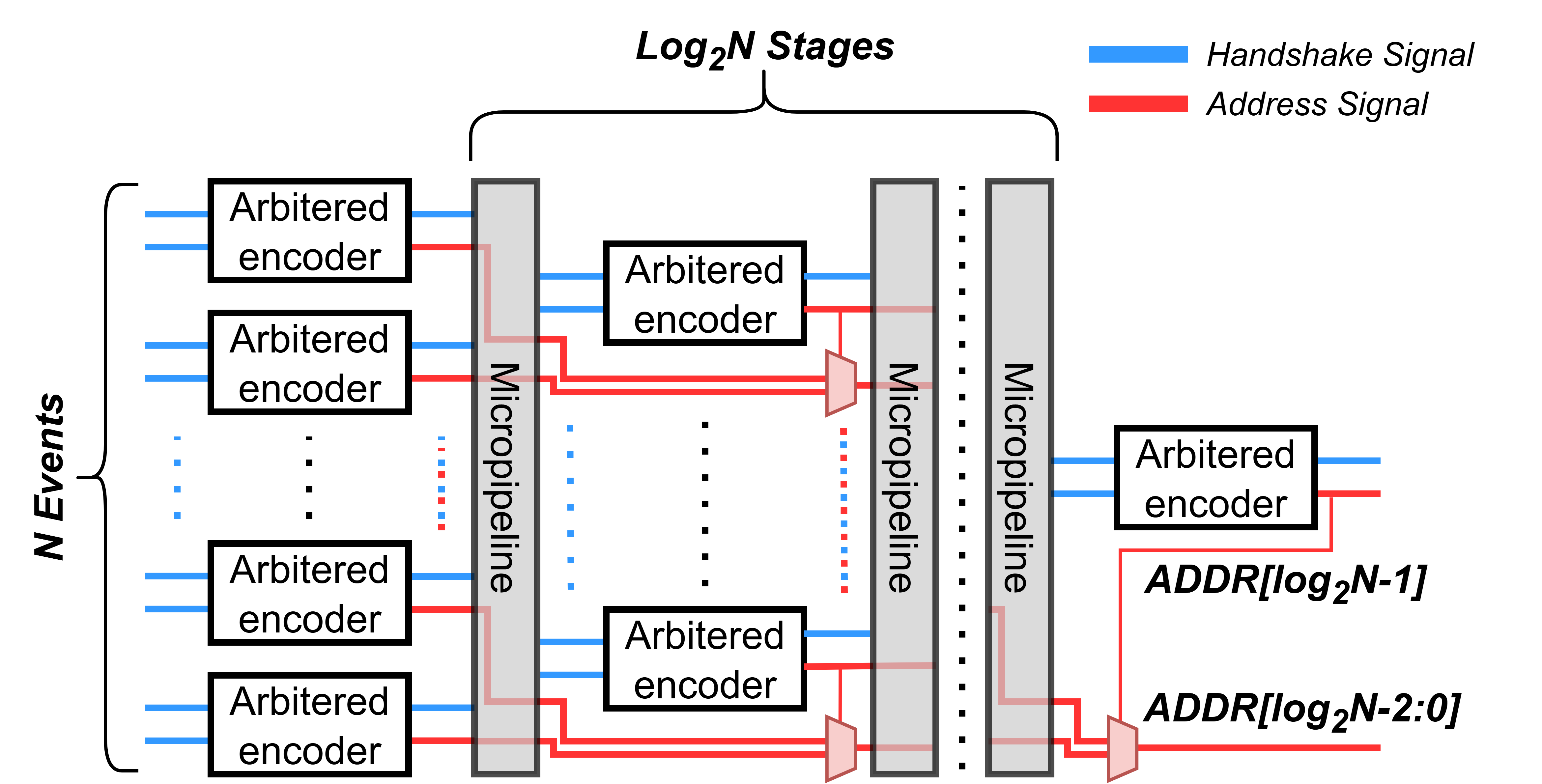}
    \caption{The hierarchical AER tree structure. Blue lines represent asynchronous handshake signals, and red lines represent address data paths. At each pipeline stage, the accumulated address from the previous stage is multiplexed with the current stage's event address, merged, and forwarded.}
    \label{fig:aer_arch}
\end{figure}

In this work, we propose an architecture that aggregates events using a hierarchical binary tree. As in Fig.~\ref{fig:aer_arch}, at each pipeline stage, the arbitered encoder receives inputs from its two child node encoders at the previous stage. The encoder generates a single-bit address that serves as a multiplexer select signal, determining which address from the previous stage is concatenated with the new bit. For an $N$-event system, this recursive multiplexing and concatenation operation propagates through $\log_2(N)$ stages, ultimately yielding the complete $\log_2(N)$-bit source address of the original event.

\subsection{4-Phase Modified Semi-Decoupled Micropipeline Design}

\begin{figure}[htbp]
    \centering
    \includegraphics[width=0.9\columnwidth]{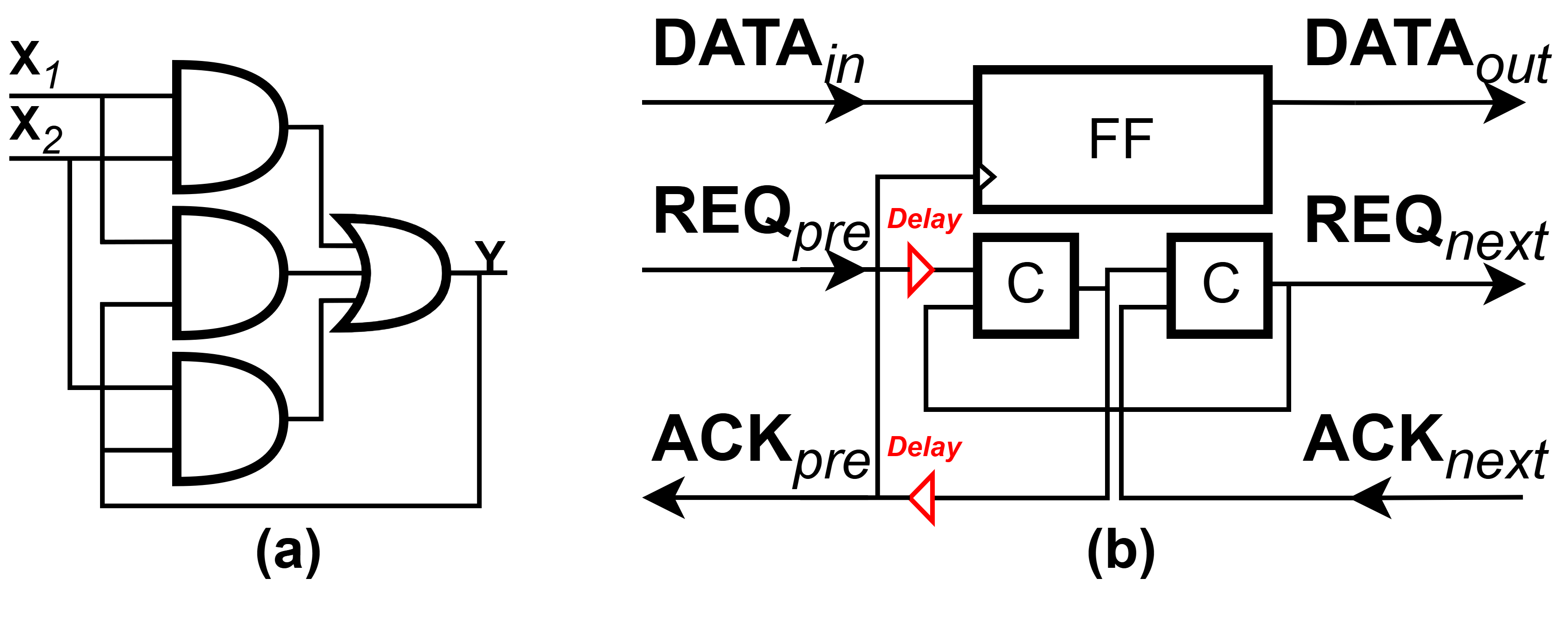}
    \caption{(a) Earle latch-based C-elements. (b) Modified 4-phase Semi-Decoupled Micropipeline architecture using C-elements with flip-flop registers. The Delay is inserted to meet the timing constraint for the combination logic between pipeline stages, which is primarily a multiplexer in our AER design.}
    \label{fig:micropipeline}
\end{figure}

As illustrated in Fig.~\ref{fig:aer_arch}, to sustain high event throughput, we employ an asynchronous 4-phase bundled-data micropipeline. While 2-phase protocols can increase throughput by capturing data on both the rising and falling edges of a handshake signal, they require specialized logic to trigger standard flip-flops (FFs)—such as the click-based protocol \cite{wu_design_2021}—or rely on transparent latches for data registration \cite{sutherland_micropipelines_1989}. These workarounds inherently introduce additional power and area overhead. In contrast, the 4-phase approach captures data exclusively on the rising edge, enabling the direct integration of standard FFs with minimal hardware overhead. It maintains high throughput by efficiently hiding the address multiplexing delay within the return-to-zero period of the handshake. To manage dense bursts of spikes and overcome the limitations of basic micropipelines \cite{sutherland_micropipelines_1989}—which must wait for an acknowledgment (ACK) from the subsequent stage before acknowledging a preceding request—the design adopts a semi-decoupled architecture \cite{furber_four-phase_1996}. This architecture utilizes compact Earle latch-based C-elements as shown in Fig.~\ref{fig:micropipeline}(a), comprising just three AND gates and one OR gate \cite{murphy_design_2012}. By decoupling the input and output handshakes, this dual C-element configuration allows the pipeline to respond to incoming requests immediately, independent of the subsequent stage's readiness.

In the proposed architecture, FFs are clocked strictly on the rising edge of the locally generated ACK signal rather than on the incoming REQ, as shown in Fig.~\ref{fig:handshake_timing}, and the circuit structure is shown in Fig.\ref{fig:micropipeline}(b). This significantly relaxes setup time requirements by hiding combinational datapath delays behind the controller's return-to-zero phase. To prevent overrun hazards and ensure safe data capture, matched delay elements are explicitly inserted into the REQ and ACK lines by Verilog parameter to encompass worst-case logic delays between stages, as illustrated in Fig.~\ref{fig:micropipeline}(b) and analyzed in Section \uppercase\expandafter{\romannumeral 3}.

\subsection{Stochastic Event Arbitration and Encoding}

\begin{figure}[htbp]
    \centering
    \includegraphics[width=0.7\columnwidth]{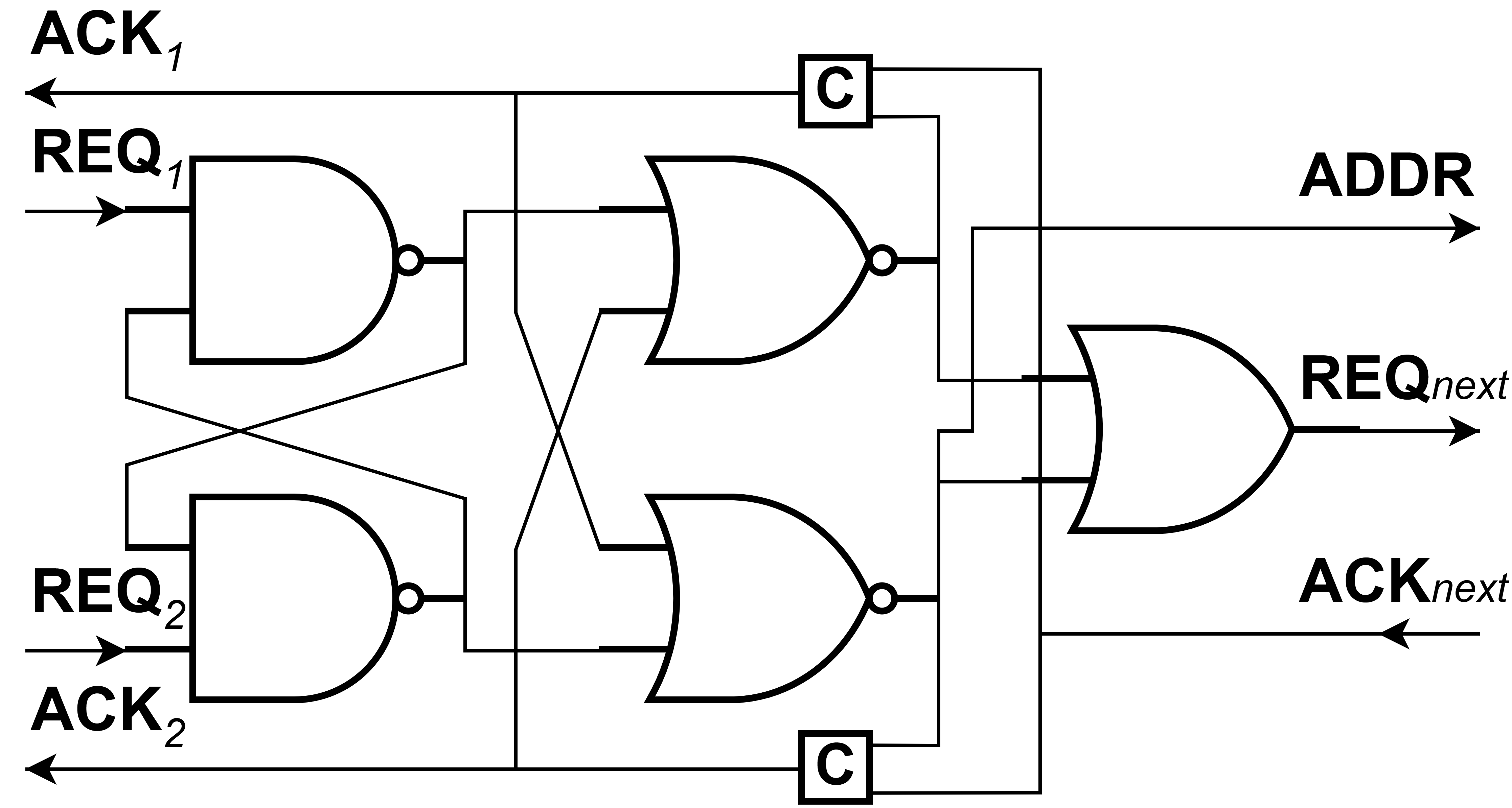}
    \caption{Multiplexed Arbitration circuit. The inherent metastability of the cross-coupled NAND gates provides stochastic resolution for simultaneous event collisions.}
    \label{fig:arbiter}
\end{figure}

To resolve inevitable event collisions in tree-based AER architectures without the latency and state-tracking overhead of deterministic rotation priority schemes \cite{wei_asynchronous_2019}, we adopt a compact Multiplexed Arbitration design based on cross-coupled NAND gates \cite{martin_asynchronous_2006} (Fig.~\ref{fig:arbiter}). During asynchronous arrivals, the faster signal drives its respective NAND output low, locking out the competing channel while asserting $\mathrm{REQ}_{\mathrm{next}}$ and the encoded address ($\mathrm{ADDR}$). The losing request is held safely pending until the downstream node clears the transaction via $\mathrm{ACK}_{\mathrm{next}}$. 

When competing events arrive near-simultaneously, the cross-coupled gates enter a metastable state, relying on inherent silicon mismatch and thermal noise to eventually break symmetry and randomly declare a winner. Rather than a hardware flaw, this stochastic tie-breaking mechanism mirrors biological neural networks, which harness cellular thermodynamic noise to prevent systemic deadlocks and enable flexible decision-making \cite{braun_stochasticity_2021}. Thus, the asynchronous arbiter leverages physical noise for biologically plausible conflict resolution while preserving the temporal sparsity of neural activity. In this work the arbiter has been implemented at the input of the AER on the FPGA. 

\section{Circuit Implementation}

\subsection{Standard-Cell Circuit Implementation Flow}
The proposed architecture is implemented using a fully automated digital RTL-to-GDSII flow. The parameterized AER module is first modeled in Verilog, beginning with the foundational Muller C-element \cite{murphy_design_2012}, which forms the core of the micropipeline (Fig.~\ref{fig:micropipeline}) and the arbitered encoder (Fig.~\ref{fig:arbiter}). These components are hierarchically instantiated to construct the complete AER routing tree (Fig.~\ref{fig:aer_arch}), utilizing only standard digital logic gates. 

The Verilog implementation is highly parameterized, allowing the number of input events, the number of pipeline stages, and the address bitwidths to be scaled according to application requirements. For this work, an 8-event AER template was implemented, fabricated, and characterized; the silicon measurement results are detailed in Section \uppercase\expandafter{\romannumeral 4}. 

The physical implementation leverages the Cadence EDA toolkit: Genus for logic synthesis, Innovus for place-and-route (PnR), and Virtuoso for final padframe integration and global routing. Behavioral, post-synthesis, and post-layout simulations were conducted using Xcelium, utilizing Standard Delay Format (SDF) back-annotation at each stage to ensure timing accuracy. However, due to the absence of a global clock in asynchronous designs, standard Static Timing Analysis (STA) tools often struggle to verify these circuits seamlessly. To address this, an automated timing closure methodology was developed and is detailed in Section III-B.

\subsection{Delay Matching and Timing Constraints}

Rigorous timing analysis is critical in bundled-data pipelines to prevent metastable states from propagating. Previous works \cite{ouyang_scalable_2024, wei_asynchronous_2019, gibiluka_bundled-data_2015} % 
achieve timing closure using the \textit{set\_min\_delay} command in Synopsys tools to constrain the handshake cycle. However, many synthesis engines, including Cadence Genus, lack equivalent support for asynchronous loops. To overcome this, we developed a tool-agnostic, automated methodology utilizing standard SDC constraints and iterative simulation.

As illustrated in Fig.~\ref{fig:handshake_timing}(b), the $\mathrm{ACK}$ signals act as local clocks. Data launched at the current stage ($\mathrm{DATA}_{\mathrm{pre, out}}$) on the rising edge of $\mathrm{ACK}_{\mathrm{pre}}$ must propagate through the combinational multiplexing logic and stabilize at the next stage ($\mathrm{DATA}_{\mathrm{next, in}}$) before the \textit{subsequent} rising edge of $\mathrm{ACK}_{\mathrm{pre}}$. Therefore, the datapath delay must strictly satisfy: $Data_{\mathrm{delay}} < T_{\mathrm{period, ack}} - T_{\mathrm{setup}}$. 

To automatically satisfy this constraint across the entire AER without relying on custom EDA features, we employ a shell-scripted iterative methodology:

\begin{enumerate}
    \item \textbf{Initial Synthesis:} Apply an initial \textit{set\_max\_delay} SDC constraint to the datapath and synthesize the gate-level netlist with a baseline number of control path delay buffers.
    \item \textbf{Timing Extraction:} Perform post-synthesis simulation with Standard Delay Format (SDF) back-annotation to extract the actual handshake cycle duration ($T_{\mathrm{period, ack}}$) for each stage.
    \item \textbf{Iterative Optimization:} Dynamically update the SDC \textit{set\_max\_delay} constraint to be strictly less than 20\% of the extracted $T_{\mathrm{period, ack}}$ to ensure robust setup margins. If the synthesis Static Timing Analysis (STA) fails this new constraint, incrementally add delay buffers to the control path to extend $T_{\mathrm{period, ack}}$ and repeat Steps 2 and 3 until the design is hazard-free.
    \item \textbf{Physical Closure:} Execute Place and Route (PnR). If interconnect parasitics introduce new post-route datapath violations, increment the control delay buffers and re-verify.
\end{enumerate}

By combining standard SDC bounds with automated log-parsing and simulation loops, this methodology provides a generic, highly reliable timing closure solution for standard-cell asynchronous designs utilizing edge-triggered flip-flops.

\section{Measurement Results}
% This is where your taped-out chip data shines.
\subsection{Test Setup and Methodology}

\begin{figure}[htbp]
    \centering
    % Remember to rename your file to avoid the double-dot issue!
    \includegraphics[width=0.65\columnwidth]{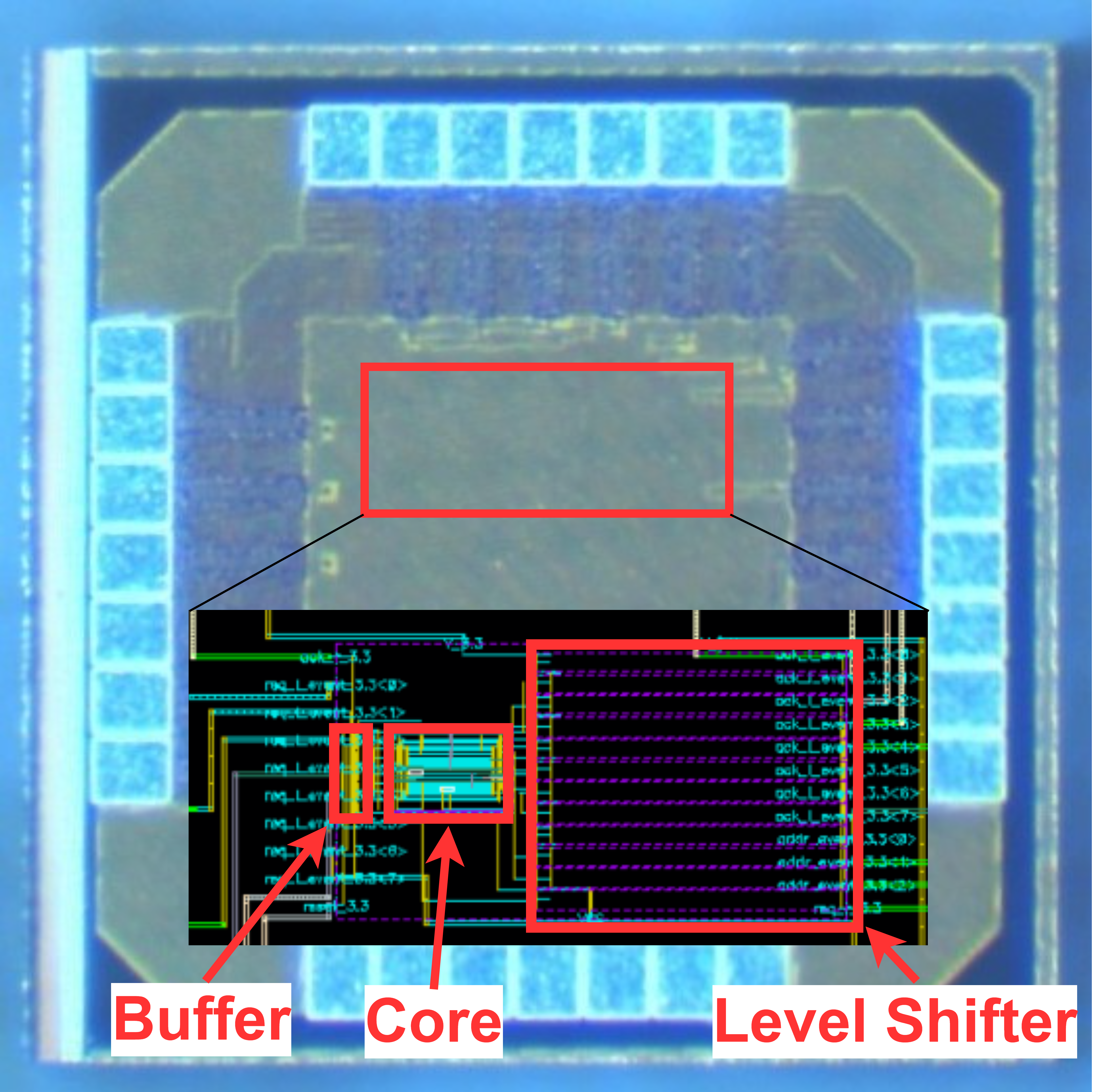}
    \caption{Die micrograph and layout details of the fabricated AER encoder in 65 nm CMOS. The active asynchronous core measures $65.57~\mu\mathrm{m} \times 44.6~\mu\mathrm{m}$.}
    \label{fig:layout}
\end{figure}

A prototype 8-event AER was fabricated using a $65~\mathrm{nm}$ CMOS process, with the physical die details illustrated in Fig.~\ref{fig:layout}. The stochastic arbiters were implemented exclusively at the input stage on the FPGA to verify their functionality. To evaluate the physical performance of the asynchronous router, the test chip was integrated into a breadboard-based test setup. A Xilinx Spartan-7 FPGA was utilized to inject high-speed, full-scan event streams into the test chip, ensuring the chip is working at full load. 

The encoded output addresses and corresponding handshake signals are continuously routed back to the FPGA for real-time functional verification. Concurrently, a 100~MHz Digilent Analog Discovery logic analyzer monitors the asynchronous control lines and output address data to validate routing correctness, assess pipeline performance, and ensure strict adherence to the bundled-data protocol. Latency and power measurements are averaged across all events captured within a 1~s measurement window. To quantify the system's energy efficiency, dynamic power consumption is calculated by measuring the voltage drop across a precision shunt resistor placed in series with the $1.2~\mathrm{V}$ core supply rail ($V_{\mathrm{DD}}$).

\subsection{Performance Measurement and Analysis}

In a tree-based micropipeline architecture, the overall system throughput is inherently bounded by the handshake completion rate of the initial pipeline stage, as depicted at the input of Fig.~\ref{fig:aer_arch}. Because the first stage is ready to process a subsequent event immediately after passing the current event forward, the maximum throughput is governed by the duration of a single handshake cycle (Fig.~\ref{fig:handshake_timing}). During silicon validation, a 100~MHz clocked FPGA was utilized to inject full-scan events into the fabricated chip. Measurements indicate an average handshake cycle duration of 30~ns between the FPGA and the asynchronous core, yielding a sustained peak throughput of $\approx$33~MEvent/s. 

System latency, measured from the rising edge of the input request (REQ) signal to the falling edge of the output REQ signal for a single event, averages 50~ns across the tree. Normalizing this delay across the pipeline depth, which is 3 stages in our prototype, provides an average propagation latency of $\approx$17~ns/(event-bit).

Because the 1.2~V supply rail is shared with a co-fabricated analog design on the same die, the total power consumption was determined by combining measured dynamic power with simulated static power. The dynamic power during event processing was measured at 4~$\mu$W utilizing the voltage drop across a precision series shunt resistor. This active measurement was carefully isolated from the aggregate baseline leakage originating from the I/O pads and the shared analog circuitry. Meanwhile, the static power dissipation of the dedicated asynchronous AER core was estimated via simulation at 4.7~$\mu$W, yielding a total power consumption of 8.7~$\mu$W. Based on this power profile, the proposed asynchronous encoder achieves an exceptionally high energy efficiency of 435~fJ per encoded event. For the 3-bit prototype design, this equates to an ultra-low 145~fJ/(event-bit).

\subsection{Performance Summary and Architecture Comparison}

Table \ref{tab:comparison} compares the proposed 65~nm AER encoder with recent asynchronous architectures \cite{qiao_bi-directional_2018, purohit_field-programmable_2022, ouyang_scalable_2024}. The design achieves an exceptional measured energy efficiency of 145~fJ/(event-bit), outperforming both the simulated custom-EDA FP-AER (102.33~pJ/(event-bit)) \cite{purohit_field-programmable_2022} and the full-custom 28~nm Bi-AER (420~fJ/(event-bit)) \cite{qiao_bi-directional_2018}. While our 33~MEvent/s measured throughput exceeds the Bi-AER, it is theoretically outperformed by the simulated mask-based topology of Ouyang et al. ($\approx$600~MEvent/s) \cite{ouyang_scalable_2024}. Similarly, the Bi-AER leverages full-custom 28~nm transistor optimization to reach a fundamentally different latency regime (0.19~ns/(event-bit)) compared to the proposed 16.67~ns/(event-bit) \cite{qiao_bi-directional_2018}. However, this deliberate trade-off for automated standard-cell synthesizability does not bottleneck practical systems. State-of-the-art CMOS artificial sensory receptors peak at spike rates of only 1~kHz \cite{subbulakshmi_radhakrishnan_biomimetic_2021}, and dense dynamic vision sensor bursts operate at the microsecond scale \cite{lichtsteiner_128times_2008}. Thus, our 50~ns total latency and 33~MEvent/s bandwidth outpace peak biological and artificial requirements by orders of magnitude, proving that superior energy efficiency and sufficient routing speeds are achievable entirely within standard commercial EDA workflows.

\begin{table}[htbp]
\centering
\caption{Measured Performance and Architecture Comparison}
\label{tab:comparison}
\resizebox{\columnwidth}{!}{%
\begin{tabular}{@{} l c c c c @{}}
\toprule
\textbf{Metric} & \textbf{This Work} & \begin{tabular}[c]{@{}c@{}}\textbf{Purohit} \\ \textbf{\cite{purohit_field-programmable_2022}}\end{tabular} & \begin{tabular}[c]{@{}c@{}}\textbf{Ouyang} \\ \textbf{\cite{ouyang_scalable_2024}}\end{tabular} & \begin{tabular}[c]{@{}c@{}}\textbf{Bi-AER} \\ \textbf{\cite{qiao_bi-directional_2018}}\end{tabular} \\ \midrule

\textbf{Architecture} & Binary Tree & Mixed Tree & \begin{tabular}[c]{@{}c@{}}Dual-Level \\ Mask\end{tabular} & Shared Bus \\[2ex]

\textbf{Implementation} & \begin{tabular}[c]{@{}c@{}}65nm CMOS \\ \textbf{Standard Cell}\end{tabular} & \begin{tabular}[c]{@{}c@{}}65nm CMOS \\ \textbf{Custom EDA}\end{tabular} & \begin{tabular}[c]{@{}c@{}}65nm CMOS \\ \textbf{Standard Cell}\end{tabular} & \begin{tabular}[c]{@{}c@{}}28nm CMOS \\ \textbf{Custom Circuit}\end{tabular} \\[2ex]

\textbf{Synchronization} & Asynchronous & Asynchronous & Asynchronous & Asynchronous \\[2ex]

\textbf{Throughput} & 33 MEvent/s & N/A & \textbf{$\approx$600 MEvent/s}\textsuperscript{*} & 28.6 MEvent/s \\[2ex]

\textbf{Latency / Event-Bit} & 16.67 ns & 20.23 ns\textsuperscript{*} & 49.88 ns\textsuperscript{*} & \textbf{0.19 ns} \\[2ex]

\textbf{Energy / Event-Bit} & \textbf{145 fJ} & 102.33 pJ\textsuperscript{*} & N/A & 420 fJ \\ \bottomrule
\multicolumn{5}{l}{\textsuperscript{*} \scriptsize Simulated results; not verified in silicon.}
\end{tabular}%
}
\end{table}

\section{Conclusion}
This paper presents a fully synthesizable, tree-based asynchronous AER encoder that integrates seamlessly into standard digital workflows. Silicon measurements from our 65 nm CMOS prototype validate the architecture's efficiency, demonstrating an average throughput of 33~MEvent/s, a 17 ns/(event-bit) propagation latency, and an ultra-low energy footprint of 435~fJ/event. This exceptional energy efficiency makes the design highly suitable for power-constrained neuromorphic edge devices, such as brain-machine interfaces and real-time neural decoders. Additionally, its modular and parameterized architecture ensures ready scalability to meet the massive event-routing demands of extremely large-scale neuromorphic systems.
% References section using your specific .bib file
\bibliographystyle{IEEEtran}
\bibliography{IEEEcontrol, AER}

\end{document}